\documentclass[11pt,a4paper,reqno]{amsart}%
\usepackage{amsthm,amsmath,amsfonts,amssymb,amsxtra,bookmark,appendix,dsfont,bm}

\theoremstyle{plain}
\newtheorem{theorem}{Theorem}

\newtheorem{proposition}[theorem]{Proposition}

\theoremstyle{definition}

\theoremstyle{remark}

\DeclareMathOperator{\Tr}{Tr}
\DeclareMathOperator{\Ker}{Ker}

\def\bq{\begin{eqnarray}}
\def\eq{\end{eqnarray}}
\def\bqq{\begin{align*}}
\def\eqq{\end{align*}}

\def\nn{\nonumber}

\def\eps{\varepsilon}

\renewcommand{\epsilon}{\varepsilon}

\def\bU{\mathbb{U}}

\def\cF {\mathcal{F}}

\def\cV {\mathcal{V}}
\def\R {\mathbb{R}}
\def\C {\mathbb{C}}
\def\N {\mathcal{N}}
\def\U {\mathbb{U}}
\def\cU {\mathcal{U}}
\def\cJ {\mathcal{J}}

\def\cA{\mathcal{A}}

\def\cK{\mathcal{K}}

\def\F {\mathcal{F}}

\def\gH{\mathfrak{h}}

\def\h{\gH}

\def\cV {\mathcal{V}}
\def\R {\mathbb{R}}
\def\C {\mathbb{C}}
\def\N {\mathcal{N}}
\def\U {\mathbb{U}}
\def\J {\mathcal{J}}
\def\cJ {\mathcal{J}}
\def\S {\mathcal{S}}

\def\cA{\mathcal{A}}

\def\d{{\rm d}}

\newcommand{\bH}{\mathbb{H}}
\newcommand{\dGamma}{{\ensuremath{\rm d}\Gamma}}

\allowdisplaybreaks

\title[Bogoliubov Hamiltonians]{Recent advances in the theory of Bogoliubov Hamiltonians}

\author[M. Napi\'orkowski]{Marcin Napi\'orkowski}
\address{Department of Mathematical Methods in Physics, Faculty of Physics, University of Warsaw,  Pasteura 5, 02-093 Warszawa, Poland}
\email{marcin.napiorkowski@fuw.edu.pl}

\begin{document}
\date{\today}

\begin{abstract} 
Bosonic quadratic Hamiltonians, often called Bogoliubov Hamiltonians, play an important role in the theory of many-boson systems where they arise in a natural way as an approximation to the full many-body problem. In this note we would like to give an overview of recent advances in the study of bosonic quadratic Hamiltonians. In particular, we relate the reported results to what can be called the time-dependent diagonalization problem.   \\
\\
\\
\\
\center{\textit{\larger{Dedicated to Herbert Spohn on the occasion of his 70th birthday}}}
\end{abstract}

\maketitle


\section{Introduction}
 Bosonic quadratic Hamiltonians are operators that are formally given by expressions of the form
\begin{equation}
\bH=\sum_{ij} h_{ij} a^*_i a_j +\frac12\sum_{ij} k_{ij} a^*_i a^*_j +\frac12\sum_{ij}\overline{k}_{ij} a_i a_j.  \label{eq:BogHamformal}
\end{equation}
Here $h_{ij}$ is a self-adjoint matrix ($h_{ij}=h^*_{ij}=\overline{h}^{\#}_{ij}$), $\overline{k}_{ij}$ is the complex conjugate of the matrix $k_{ij}$ and 
 $a^*_i/a_i$ are the bosonic creation/annihilation operators on the Fock space satisfying the canonical commutation relations
$$[a^*_i,a^*_j]=[a_i,a_j]=0, \qquad [a_i, a^*_j]=\delta_{ij}.$$
If the sums in \eqref{eq:BogHamformal} are finite, which corresponds to the fact that the underlying one-particle Hilbert space is finite dimensional, then the operator $\bH$ is well-defined. However, if the one-particle Hilbert space is infinite dimensional, then the analysis of the operator \eqref{eq:BogHamformal} becomes more complicated. This includes even a proper definition of the formal expression given above. The goal of this note is to review recent results on $\bH$ in the infinite dimensional case.

Bogoliubov Hamiltonians are used in many different situations. In the context of quantum field theory, for example, they appear as quantization of a scalar field with a mass-like perturbation \cite{Der-14}. In the context of many-body quantum mechanics, quadratic Hamiltonians arise in a natural way as an approximation to the full many-body problem. In fact, the name \textit{Bogoliubov Hamiltonian} is related to such an approximation performed by Nikolay Bogoliubov in the famous 1947 paper \textit{On the theory of superfluidity} \cite{Bog-47}. A system of $N$ bosons (confined in a box of size $V=L^3$ with periodic boundary conditions) that interact through a two-body potential $v$ is described by the second quantized many-body Hamiltonian
\begin{gather}
H=\sum_{p\in \frac{2\pi}{L}\mathbb{Z}^3} p^2 a_p^*a_p+\frac{1}{2L^3}\sum_{p,q,k\in \frac{2\pi}{L}\mathbb{Z}^3}\hat{v}(k)a_{p+k}^* a_{q-k}^* a_q a_p.  \label{eq:manybodyHam}
\end{gather} 
 As explained in detail in \cite{ZagBru-01}, Bogoliubov argued that at low energies this system can be effectively described by a quadratic Hamiltonian 
\begin{equation}
\bH_{\text{Bog}}=\frac{N^2}{2V} +\sum_{p\neq 0}p^2 a^*_p a_p+\frac{N}{2V}\sum_{p\neq 0}\hat{v}(p)(2 a^*_p a_p+a^*_p a^*_{-p}+a_p a_{-p}). \label{eq:quadrBog47}
\end{equation}
Finally, Bogoliubov noticed that introducing operators 
\begin{equation}
b_p =c_{p}a_p+s_{p}a^*_{-p},  \label{eq:quasiparticledefinition}
\end{equation}
with some appropriately chosen real $c_p$ and $s_p$, one can \textit{diagonalize} $\bH_{\text{Bog}}$, that is rewrite it in the form
\begin{equation}
\bH_{\text{Bog}}=E+\sum_p e(p) b_p^* b_p  \label{eq:OriginalBogHamDiag}
\end{equation}
where the constant $E$ and function $e(p)$ depend on the parameters of the system (in particular they depend on the interaction potential $v$). From \eqref{eq:OriginalBogHamDiag} one can easily determine the spectrum of $\bH_{\text{Bog}}$. Also, \eqref{eq:OriginalBogHamDiag} provides a basis for the interpretation that at small energies the collective behaviour of many particles can be interpreted as a system of non-interacting \textit{quasiparticles} (\cite{DerMeiNap-13,Napiorkowski-14}).

The procedure that was applied by Bogoliubov to obtain \eqref{eq:quadrBog47} from \eqref{eq:manybodyHam} has since then been formulated in a more abstract setting and has become a standard tool in different subfields of condensed matter physics \cite{DerNapSol-13}. A rigorous justification of this approximation, which is, for example, proving that the spectrum of $\bH_{\text{Bog}}$ indeed approximates the spectrum of $H$, has only been achieved in some situations \cite{Seiringer-11, LewNamSerSol-15,GreSei-13,DerNap-13,NamSei-15, BBCS-17,BBCS-18}.

The other important idea introduced by Bogoliubov, which is the diagonalization of \eqref{eq:quadrBog47} leading to \eqref{eq:OriginalBogHamDiag}, is what we want to address in this review. Using the so-called Lie formula, it is easy to see that \eqref{eq:quasiparticledefinition} can be written as
$$b_p =c_{p}a_p+s_{p}a^*_{-p}= e^{-X}a_p e^{X}$$
with $X=\sum_{p\neq 0}\beta_p (a^*_p a^*_{-p}-a_p a_{-p})$ for some appropriately chosen $\beta_p$. Since $e^X$ is unitary, we see that \eqref{eq:OriginalBogHamDiag} has been obtained from \eqref{eq:quadrBog47} by a unitary transformation on the Fock space. In particular, the $b_p$ operators also obey the canonical commutation relations. The obvious questions arise: can this be done for any quadratic bosonic Hamiltonian? If not, then under which conditions?

Before we can review the answers to these questions we need to formulate them in a rigorous setting. This will be done in the next section. In Section \ref{sec:proof-remarks} we will sketch the main ideas behind the proofs. These parts of this review are based on \cite{NamNapSol-16}. In Section \ref{sec:qf-dynamics} we will mention a result about time-dependent quadratic Hamiltonians. Finally, in Section \ref{sec:time-dep-diag}, we will make a connection between the previous two sections in addressing the issue of what can be called a time-dependent diagonalization problem.

 \medskip

\noindent{\bf Acknowledgments.} The support of the National Science Centre (NCN) project Nr. 2016/21/D/ST1/02430 is gratefully acknowledged. 
 
\section{Main result}
\subsection{Fock space formalism.} 
Let us introduce the mathematical setting. Our one-body Hilbert space $\h$ is a complex separable Hilbert space with inner product $\langle.,.\rangle$ which is linear in the second variable and anti-linear in the first. The bosonic Fock space  is defined by
$$\F(\h):=\bigoplus_{N=0}^\infty \bigotimes_{\text{sym}}^N \h = \C \oplus \gH \oplus \left(\gH\otimes_s \gH \right) \oplus \cdots$$
Let $h>0$ be a self-adjoint operator on $\gH$. Then it can be lifted to the Fock space by
$$
\dGamma(h):= \bigoplus_{N=0}^\infty (\sum_{j=1}^N h_j) =  0 \oplus h \oplus (h\otimes 1 + 1 \otimes h) \oplus \cdots  
$$
A typical example is that $h=-\Delta+V(x)$ on $\gH=L^2(\R^d)$, where $V$ is an external potential which serves to bind the particles. Another example is given by the identity operator. Then $\dGamma(1)$ is the particle number operator denoted by $\N$.  The operator $\dGamma(h)$ is well-defined on the core 
$$
\bigcup_{M\ge 0} \bigoplus_{n=0}^M  \bigotimes_{\rm sym}^n D(h)
$$
and it can be extended to a positive self-adjoint operator on Fock space by Friedrichs' extension. 

To describe quadratic Hamiltonians on bosonic Fock space, we introduce the {\em creation} and {\em annihilation} operators. For any vector $f \in \gH$, the creation operator $a^*(f)$ and the annihilation operator $a(f)$ are defined by the following actions  
\begin{align*} 
{a^*}({f_{N+1}})\left( {\sum\limits_{\sigma  \in {S_{N}}} {{f_{\sigma (1)}} } \otimes ... \otimes {f_{\sigma (N)}}} \right) = \frac{1}{\sqrt {N+1}} \sum\limits_{\sigma  \in {S_{N+1}}} {{f_{\sigma (1)}} }  \otimes ... \otimes {f_{\sigma (N+1)}},
\end{align*}
\begin{align*} 
  a(f_{N+1})\left( {\sum\limits_{\sigma  \in {S_N}} {{f_{\sigma (1)}} } \otimes ... \otimes {f_{\sigma (N)}}} \right) = \sqrt N \sum\limits_{\sigma  \in {S_N}} {\left\langle {f_{N+1},{f_{\sigma (1)}}} \right\rangle } {f_{\sigma (2)}} \otimes ... \otimes {f_{\sigma (N)}},
\end{align*}
for all $f_1,...,f_{N+1}$ in $\gH$, and all $N=0,1,2,...$. These operators satisfy the canonical commutation relations (CCR)
\bq \label{eq:CCR}
[a(f),a(g)]=0, \quad [a^*(f),a^*(g)]=0, \quad [a(f),a^*(g)]=\langle f, g\rangle, \quad \forall f,g\in \gH.
\eq
In particular, for every $f \in \gH$ we have 
\bq
a(f)|0\rangle =0 \nn 
\eq
where $|0\rangle=1\oplus 0 \oplus 0 \cdots$ is the Fock space vacuum. If $\{f_n\}_{n\ge 1}\subset D(h)$ is an arbitrary orthonormal basis for $\gH$, then it is common to denote the operators $a^*(f_i)$ and $a(f_i)$ by $a^*_i$ and $a_i$ respectively. This explains the notation used in the introduction. In particular, the operators $a_p$ used in the presentation of Bogoliubov's approach correspond to $a(V^{-1/2} e^{-ipx})$ (plane waves basis).   
\subsection{The Hamiltonian.} In general, a quadratic Hamiltonian on Fock space is a linear operator which is quadratic in terms of creation and annihilation operators. For example, $\dGamma(h)$ is a quadratic Hamiltonian because we can write
\begin{align*}
\dGamma(h)= \sum_{m,n\ge 1}\langle f_m, h f_n \rangle a^*(f_m) a(f_n) 
\end{align*}
(the sum on the right side is independent of the choice of the basis). Thus the matrix $h_{ij}$ in \eqref{eq:BogHamformal} is given by $h_{ij}=\langle f_i, h f_j\rangle$. In this paper, we will consider a general quadratic operator of the form
\bq \label{eq:quadratic-Hamiltonian}
\bH = \dGamma(h) + \frac{1}{2}\sum_{m,n\ge 1} \left( \langle J^*kf_m,f_n \rangle a(f_m) a(f_n) + \overline{\langle J^*kf_m,f_n \rangle} a^*(f_m)a^*(f_n)  \right)
\eq
where $k:\gH\to \gH^*$ is an unbounded linear operator with $D(h)\subset D(k)$ (called {\em pairing operator}) 
and $J:\gH\to \gH^*$ is the anti-unitary operator \footnote{If $C:\gH\to \mathfrak{K}$ is anti-linear, then $C^*: \mathfrak{K} \to \gH$ is defined by $\langle C^*g,f \rangle_{\gH} = \langle C f, g \rangle_{\mathfrak{K}}$ for all $f\in \gH,g\in \mathfrak{K}.$ The anti-linear map $C$ is an anti-unitary if $C^*C=1_{\gH}$ and $CC^*=1_{\mathfrak{K}}$.} defined by
$$J(f)(g)=\langle f,g \rangle,\quad \forall f,g\in \gH.$$  
Since $\bH$ remains the same when $k$ is replaced by $(k+J k^* J)/2$, we will always assume without loss of generality that 
\bq \label{eq:symmetry-b-2}
k^* = J^* k J^*.
\eq

In fact, the formula \eqref{eq:quadratic-Hamiltonian} is formal but $\bH$ can be defined properly as a quadratic form as follows. For every normalized vector $\Psi \in \cF(\gH)$ with finite particle number expectation, namely $\langle \Psi, \mathcal{N} \Psi \rangle <\infty$, its one-particle density matrices $\gamma_\Psi: \h\to \h$ and $\alpha_\Psi:\h \to \h^*$ are linear operators defined by
\bq \label{eq:def-gamma-alpha}
\left\langle {f,{\gamma _\Psi }g} \right\rangle  = \left\langle \Psi, {{a^*}(g)a(f)} \Psi \right\rangle,~~ \left\langle {Jf, \alpha _\Psi  g } \right\rangle  = \left\langle \Psi, {a^*(g)a^*(f)} \Psi\right\rangle, \quad\forall f,g\in \h.
\eq
A formal calculation using \eqref{eq:quadratic-Hamiltonian} leads to the expression 
\bq \label{eq:def-H-gamma-alpha}
\langle \Psi, \bH \Psi \rangle = \Tr(h^{1/2} \gamma_\Psi h^{1/2})+ \Re \Tr(k^* \alpha_\Psi).
\eq
The formula \eqref{eq:def-H-gamma-alpha} makes sense when $h^{1/2} \gamma_\Psi h^{1/2}$ and $k^* \alpha_\Psi$ are trace class operators. We will use \eqref{eq:def-H-gamma-alpha} to define $\bH$ as a quadratic form with a dense form domain described below.

Since $\gH$ is separable and $D(h)$ is dense in $\gH$, we can choose finite-dimensional subspaces $\{Q_n\}_{n=1}^\infty$ such that
$$ Q_1 \subset Q_2 \subset ... \subset D(h) \quad \text{and}\quad \overline{\bigcup_{n\ge 1} Q_n} = \gH.$$
Then it is straightforward to verify that  
\begin{align} \label{eq:def-mcQ} \mathcal{Q}:=\bigcup_{M\ge 0} \bigcup_{n\ge 1} \Big( \bigotimes_{\rm sym}^M Q_n \Big) 
\end{align}
is a dense subspace of $\cF(\gH)$. Moreover, for every normalized vector $\Psi \in \mathcal{Q}$, $\gamma_\Psi$ and $\alpha_\Psi$ are finite rank operators with ranges in $D(h)$ and $J D(h)$, respectively. Note that $JD(h)\subset D(k^*)$ because $D(h)\subset D(k)$ and $k^*=J^*kJ^*$. Thus $h^{1/2} \gamma_\Psi h^{1/2}$ and $k^* \alpha_\Psi$ are trace class and $\langle \Psi, \bH \Psi \rangle$ is well-defined by \eqref{eq:def-H-gamma-alpha} for every normalized vector $\Psi$ in $\mathcal{Q}$.

The first result we would like to report is the following \cite[Lemma 9]{NamNapSol-16}

\begin{proposition} \label{lem:bounded-below-H}  Assume that $h>0$, $k^*=J^*kJ^*$,  $kh^{-1}k^* \le JhJ^*$ and $\Tr(kh^{-1}k^*)<\infty$. Then  
$$\bH \ge -\frac{1}{2} \Tr(kh^{-1}k^*)$$
as a quadratic form. Moreover, if $kh^{-1}k^* \le \delta JhJ^*$ for some $0\le \delta<1$, then 
$$ (1+\sqrt{\delta}) \dGamma(h) + \frac{\sqrt{\delta}}{2} \Tr( k h^{-1} k^*) \ge \bH \ge (1+\sqrt{\delta}) \dGamma(h) - \frac{\sqrt{\delta}}{2} \Tr( k h^{-1} k^*)  $$
as quadratic forms. Consequently, the quadratic form $\bH$ defines a self-adjoint operator,  still denoted by $\bH$, such that $\inf \sigma(\bH) \ge -\frac{1}{2} \Tr(kh^{-1}k^*)$.
\end{proposition} 

Thus, the above proposition provides conditions under which the formal expressions \eqref{eq:BogHamformal}/\eqref{eq:quadratic-Hamiltonian} make sense. In the context of the earlier results (which will be reviewed after the statement of the main theorem) the main improvement is that we do not have to assume that $k$ is a bounded operator.

\subsection{Diagonalization.} As mentioned in the introduction, a key feature of the quadratic Hamiltonians in Bogoliubov's theory is that they can be diagonalized to those of {non-interacting} systems by a special class of unitary operators which preserve the CCR algebra. It turns out that the diagonalization problem on Fock space can be associated to a diagonalization problem on $\gH\oplus \gH^*$ in a very natural way.

Since we will consider transformations on $\gH\oplus \gH^*$, it is convenient to introduce the {generalized annihilation} and {creation} operators 
\bq \label{eq:def-A=a+a*}
A(f\oplus Jg)=a(f)+a^*(g), \quad A^*(f\oplus Jg)=a^*(f)+a(g),\quad \forall f,g\in \h.
\eq
They satisfy the conjugate and canonical commutation relations 
\bq \label{eq:conjugate-CCR}
A^*( F_1)=A(\J F_1),\quad \left[ {A(F_1 ),A^* (F_2 )} \right] = (F_1 ,\S F_2 ),\quad \forall F_1,F_2\in  \h\oplus \h^*
\eq 
where we have introduced the block operators on $\gH\oplus \gH^*$
\bq
\S = \left( \begin{gathered}
  1~{\text ~~~}0 \hfill \\
  0~~-1 \hfill \\ 
\end{gathered}  \right),\quad \J = \left( {\begin{array}{*{20}c}
   0 & {J^* }  \\
   J & 0  \\
 \end{array} } \right). \label{eq:def-S-J}
\eq
Note that $S=S^{-1}=S^*$ is a unitary on $\gH\oplus \gH^*$ and $\cJ=\cJ^{-1}=\cJ^{*}$ is an anti-unitary. 

We say that a bounded operator $\cV$ on $\gH\oplus \gH^*$ is {\em unitarily implemented} by a unitary operator $\bU_\cV$ on Fock space if 
\bq \label{eq:V-Uv-action}
\bU_\cV A(F) \bU_\cV^* = A(\cV F), \quad \forall F\in \gH\oplus \gH^*.
\eq  
It is easy to see that if \eqref{eq:V-Uv-action} holds true, then the CCR \eqref{eq:conjugate-CCR} imply the following compatibility conditions 
\begin{align}\label{eq:Bogoliubov-transformation}\cJ \cV\cJ=\cV, \quad \cV^* S \cV = S = \cV S \cV^*.
\end{align}
Any {bounded} operator $\cV$ on $\gH\oplus \gH^*$ satisfying \eqref{eq:Bogoliubov-transformation} is called a {\em Bogoliubov transformation}.

The condition $\cJ \cV \cJ =\cV$ means that $\cV$ has the block form
\bq \label{eq:u-v-block}
\cV = \left( {\begin{array}{c c}
  U & J^* V J^*  \\ 
  V & J U J^* 
\end{array}} \right)
\eq
where $U:\gH \to \gH$ and $V:\gH \to \gH^*$ are linear bounded operators. Under this form, the condition $\cV^* S \cV = S = \cV S \cV^*$ is equivalent to 
\bq \label{eq:relation-U-V}
U^* U = 1+ V^*V, \quad UU^*= 1+ J^* VV^* J, \quad  V^* J U =  U^*J^*V.
\eq
It is a fundamental result that a Bogoliubov transformation $\cV$ of the form \eqref{eq:u-v-block}  is unitarily implementable if and only if it satisfies {\em Shale's condition} \cite{Shale-62} 
\bq \label{eq:Shale}
\|V\|_{\rm HS}^2 = \Tr(V^*V)<\infty.
\eq 

Now we come back to the problem of diagonalizing $\bH$. Using the formal formula \eqref{eq:quadratic-Hamiltonian} and the assumption $k^*=J^*kJ^*$, we can write 
\bq \label{eq:quadratic-Hamiltonian-def2}
\bH= \bH_{\cA} - \frac{1}{2}\Tr(h) 
\eq
where
\begin{align} \label{eq:def-A-block}
\cA := \left( {\begin{array}{*{20}{c}}
   h & k^* \\ 
  k & J h J^* 
\end{array}} \right) 
\end{align}
and 
\begin{equation} \label{eq:Wick-quant}
\bH_{\cA}:=\frac{1}{2}\sum_{m,n\ge 1} \langle F_m, \cA F_n \rangle A^*(F_m)A(F_n).
\end{equation}
Here $\{F_n\}_{n\ge 1}$ is an orthonormal basis for $\gH\oplus \gH^*$ and the definition $\bH_{\cA}$ is  independent of the choice of the basis. In fact, $\bH_{\cA}$ is the Weyl quantization of $\cA$. We refer to \cite{Der-17} for more details on this aspect of quadratic Hamiltonians. Finally, note that $\cJ\cA\cJ=\cA$ because of the symmetry condition $k^*=J^*kJ^*$. 

Now let $\cV$ be a Bogoliubov transformation on $\gH\oplus \gH^*$ which is implemented by a unitary operator $\bU_\cV$ on Fock space as in \eqref{eq:V-Uv-action}. Then we can verify that $\bU_\cV \bH_{\cA} \bU_\cV^* =  \bH_{\cV \cA \cV^*}$ and hence \eqref{eq:quadratic-Hamiltonian-def2} is equivalent to 
\bq \label{eq:quadratic-Hamiltonian-def3}
\bU_{\cV}\bH \bU_{\cV}^* = \bH_{\cV\cA \cV^*} - \frac{1}{2}\Tr(h). 
\eq
In particular, if $\cV \cA \cV^*$ is {\em block diagonal}, namely
$$ 
\cV \cA \cV^* = \left( {\begin{array}{*{20}{c}}
  \xi & 0  \\ 
  0& J \xi J^* 
\end{array}} \right)
$$
for some operator $\xi:\gH\to \gH$, then \eqref{eq:quadratic-Hamiltonian-def3} reduces to 
\bq \label{eq:quadratic-Hamiltonian-def4}
\bU_{\cV}\bH \bU_{\cV}^* = \dGamma(\xi)+ \frac{1}{2} \Tr( \xi -h). 
\eq
Note that all formulas \eqref{eq:quadratic-Hamiltonian-def2}, \eqref{eq:quadratic-Hamiltonian-def3} and \eqref{eq:quadratic-Hamiltonian-def4} are formal because $h$, $\xi$ and $\xi-h$ may be not trace class. Nevertheless, the above heuristic argument suggests that the diagonalization problem on $\bH$ can be reduced to the diagonalization problem on $\cA$ by Bogoliubov transformations. This is summarized in the following two theorems \cite[Theorems 1 and 2]{NamNapSol-16}.

\begin{theorem}[Diagonalization of bosonic block operators]\label{thm:diag}\text{}\smallskip\\
\noindent {\rm (i) (Existence)}. Let $h:\gH\to \gH$ and $k:\gH\to \gH^*$ be (unbounded) linear operators satisfying $h=h^*>0$, $k^*=J^*kJ^*$ and $D(h)\subset D(k)$. Assume that the operator  $G:=h^{-1/2}J^*kh^{-1/2}$ is densely defined and extends to a bounded operator satisfying $\|G\|<1$. Then we can define the self-adjoint operator
$$
\cA := \left( {\begin{array}{*{20}{c}}
   h & k^* \\ 
  k & J h J^* 
\end{array}} \right)>0 \quad \text{on}~\gH\oplus \gH^*
$$
by Friedrichs' extension. This operator can be diagonalized by a bosonic Bogoliubov transformation $\cV$ on $\gH\oplus \gH^*$ in the sense that
\begin{align*}
\cV \cA \cV^* = \left( {\begin{array}{*{20}{c}}
  \xi & 0  \\ 
  0& J \xi J^* 
\end{array}} \right)
\end{align*}
for a self-adjoint operator $\xi>0$ on $\gH$. Moreover, we have
\bq \label{eq:V-norm-main-thm}
 \|\cV\| \le \left(\frac{1+\|G\|}{1-\|G\|} \right)^{1/4}.
 \eq
\smallskip

\noindent {\rm (ii) (Implementability)}. Assume further that $G$ is Hilbert-Schmidt. Then $\cV$ is unitarily implementable and, under the block form \eqref{eq:u-v-block},
\begin{align} \label{eq:V*V-HS}
\|V\|_{\rm HS} \le \frac{2}{1-\|G\|} \|G\|_{\rm HS}.
\end{align}
\end{theorem}

Next, we consider the diagonalization of quadratic Hamiltonians. 

\begin{theorem} [Diagonalization of quadratic Hamiltonians]\label{thm:diag-Hamiltonian} We keep all assumptions in {\rm Theorem \ref{thm:diag}} (that $\|G\|<1$ and $G$ is Hilbert-Schmidt) and assume further that $kh^{-1/2}$ is Hilbert-Schmidt. Then the quadratic Hamiltonian $\bH$, defined as a quadratic form by \eqref{eq:def-H-gamma-alpha}, is bounded from below and closable, and hence its closure defines a self-adjoint operator which we still denote by $\bH$. Moreover, if $\U_\cV$ is the unitary operator on Fock space implementing the Bogoliubov transformation $\cV$ in Theorem \ref{thm:diag}, then 
\bq \label{eq:diag-H-dGxi}
\bU_\cV \bH \bU_\cV^* = \dGamma(\xi)+ \inf\sigma(\bH).
\eq
Finally, $\bH$ has a unique ground state $\Psi_0=\bU_\cV^*|0\rangle$ whose one-particle density matrices are $\gamma_{\Psi_0}=V^*V$ and $\alpha_{\Psi_0}=JU^*J^*V$ and
\bq \label{eq:inf-H-main-thm}
\inf\sigma(\bH)=\Tr(h^{1/2}\gamma_{\Psi_0} h^{1/2})+\Re \Tr(k^*\alpha_{\Psi_0}) \ge -\frac{1}{2} \| kh^{-1/2}\|_{\rm HS}^2 .
\eq
In particular, $h^{1/2}\gamma_{\Psi_0} h^{1/2}$ and $k^* \alpha_{\Psi_0}$ are trace class. 
\end{theorem}

\subsection{Optimality of the diagonalization conditions.}
Having in mind the conditions in the statements above, let us consider the following example. Let $h$ and $k$ be multiplication operators on $\gH=L^2(\Omega,\C)$, for some measure space $\Omega$. Then $J$ is simply complex conjugation and we can identify $\gH^*=\gH$ for simplicity. Assume that $h>0$, but $k$ is not necessarily real-valued. Then 
$$
\cA := \left( {\begin{array}{*{20}{c}}
   h & k \\ 
  k & h 
\end{array}} \right)>0 \quad \text{on}~\gH\oplus \gH^*
$$
if and only if $-1<G<1$ with $G:=|k|h^{-1}$. In this case, $\cA$ is diagonalized by the linear operator
$$
\cV :=  \sqrt{\frac{1}{2}+\frac{1}{2\sqrt{1-G^2}}} \left( {\begin{array}{*{20}{c}}
  1 & \frac{-G}{1+\sqrt{1-G^2}}  \\ 
  \frac{-G}{1+\sqrt{1-G^2}}  & 1 
\end{array}} \right)
$$
in the sense that 
$$
\cV \cA \cV^* =  \left( {\begin{array}{*{20}{c}}
   \xi & 0 \\ 
  0 & \xi 
\end{array}} \right)\quad \text{with} \quad \xi:=h \sqrt{1-G^2}=\sqrt{h^2-k^2} >0. 
$$
It is straightforward to verify that $\cV$ always satisfies the compatibility conditions \eqref{eq:Bogoliubov-transformation}. Moreover, $\cV$ is bounded (and hence a Bogoliubov transformation) if and only if $\|G\|=\|kh^{-1}\|<1$  and in this case
\bq  \label{eq:comm-example-norm-V}
\|\cV\| \sim (1- \|G\|)^{-1/4}
\eq
(which means that the ratio between $\|\cV\|$ and $(1- \|G\|)^{-1/4}$ is bounded from above and below by universal positive constants). By Shale's condition \eqref{eq:Shale}, $\cV$ is unitarily implementable if and only if $kh^{-1}$ is Hilbert-Schmidt and in this case, under the conventional form \eqref{eq:u-v-block}, 
\bq  \label{eq:comm-example-HS-V}
\|V\|_{\rm HS} \sim (1- \|G\|)^{-1/4} \|G\|_{\rm HS}.
\eq
Finally, from \eqref{eq:quadratic-Hamiltonian-def4} and the simple estimates 
$$
-\frac{1}{2} k^2 h^{-1} \ge \xi -h = \sqrt{h^2-k^2} -h \ge - k^2 h^{-1}
$$
we deduce that $\bH$ is bounded from below if and only if $k h^{-1/2}$ is Hilbert-Schmidt and in this case
\bq  \label{eq:comm-example-lwb-bH}
\inf \sigma(\bH) \sim - \|k h^{-1/2}\|_{\rm HS}^2.
\eq
  
Thus, in summary, in the above commutative example we have the following {\em optimal} conditions:
\begin{itemize}

\item[$\bullet$] $\cA$ is diagonalized by a Bogoliubov transformation $\cV$ if and only if $\|kh^{-1}\|<1$;  

\item[$\bullet$] $\cV$ is unitarily implementable if and only if  $kh^{-1}$ is Hilbert-Schmidt;

\item[$\bullet$] $\bH$ is bounded from below if and only if $kh^{-1/2}$ is Hilbert-Schmidt.
\end{itemize}

If we now compare this with the conditions in Theorems \ref{thm:diag} and \ref{thm:diag-Hamiltonian}, then condition $\|G\|<1$ can be interpreted as a non-commutative analogue of the bound $\|kh^{-1}\|<1$ in the commutative case. Furthermore, the implementability condition $\|G\|_{\rm HS}<\infty$  can be interpreted as a non-commutative analogue of the condition $\|kh^{-1}\|_{\rm HS}<\infty$ in the commutative case. Finally, the condition $\|kh^{-1/2}\|_{\rm HS}<\infty$ is the same as in the commutative case. This necessary condition was proved by Bruneau and Derezi\'nski in \cite{BruDer-07} when $k$ is Hilbert-Schmidt. Note that in order to ensure that $\bH$ is bounded from below, we do not really need the conditions $\|G\|<1$ and $\|G\|_{\rm HS}<\infty$ in Theorem \ref{thm:diag}, as stated in Proposition \ref{lem:bounded-below-H}.

\subsection{Comparison with existing results}
Let us make some historical remarks. The physical model in Bogoliubov's 1947 paper \cite{Bog-47} and described in the introduction corresponds to the case when $\dim \gH=2$ and $\cA$ is a $2\times 2$ real matrix which can be diagonalized explicitly (more precisely, in his case particles only come in pairs with momenta $\pm p$ and each pair can be diagonalized independently). In fact, when $\dim \gH$ is finite, the  diagonalization of $\cA$ by symplectic matrices can be done by Williamson's Theorem \cite{Williamson-36}. We refer to H\"ormander \cite{Hormander-95} for a complete discussion on the diagonalization problem in the finite dimensional case.

In the 1950's and 1960's, Friedrichs \cite{Friedrichs-53} and Berezin \cite{Berezin-66} gave general diagonalization results in the case $\dim \gH=+\infty$, assuming that $h$ is bounded, $k$ is Hilbert-Schmidt, and $\cA\ge \mu>0$ for a constant $\mu$. Note that the gap condition $\cA\ge \mu$ requires that $h\ge \mu>0$.

In the present paper we always assume that $\cA>0$ but we do not require a gap. In some cases, the weaker assumption $\cA\ge 0$ might be also considered, but is is usually transferred back to the strict case $\cA>0$ by using an appropriate decomposition; see Kato and Mugibayashi \cite{KatMug-67} for a further discussion.

In many physical applications, it is important to consider unbounded operators. In the recent works on the excitation spectrum of interacting Bose gases, the diagonalization problem has been studied by Grech and Seiringer in \cite{GreSei-13} when $h$ is a positive operator with compact resolvent and $k$ is Hilbert-Schmidt, and then by Lewin, Nam, Serfaty and Solovej \cite[Appendix A]{LewNamSerSol-15} when $h$ is a general unbounded operator satisfying $h\ge \mu>0$. 


Very recently, in 2014, Bach and Bru \cite{BacBru-15} established for the first time the diagonalization problem when $h$ is not bounded below away from zero. They assumed that $h>0$, $\|kh^{-1}\|<1$ and $kh^{-s}$ is Hilbert-Schmidt for all $s\in [0,1+\eps]$ for some $\eps>0$ (see conditions (A2) and (A5) in \cite{BacBru-15}). 

\section{Remarks about the proof}\label{sec:proof-remarks}
Let us now make a few remarks about the proofs. All details can be found in \cite{NamNapSol-16}.

The obtain the first bound in Proposition \ref{lem:bounded-below-H} we notice that since $\gamma_\Psi$ and $\alpha_\Psi$ are finite-rank operators (when $\Psi$ is a normalized vector in the domain $\mathcal{Q}$), we can use the cyclicity of the trace and the Cauchy-Schwarz inequality to write
\begin{align*} 
|\Tr( k^* \alpha_\Psi )| &= |\Tr( \alpha_\Psi k^*)| = |\Tr((1+J\gamma_\Psi J^*)^{-1/2}\alpha_\Psi h^{1/2}h^{-1/2} k^* (1+J\gamma_\Psi J^*)^{1/2})| \nn \\
&\le \|(1+J\gamma_\Psi J^*)^{-1/2}\alpha_\Psi h^{1/2}\|_{\rm HS} \| h^{-1/2} k^* (1+J\gamma_\Psi J^*)^{1/2} \|_{\rm HS} \nn \\
&= \left[ \Tr( h^{1/2}  \alpha_\Psi^* (1+J\gamma_\Psi J^*)^{-1}\alpha_\Psi  h^{1/2} ) \right]^{1/2} \nn \\
&\quad \times \left[ \Tr \Big( (1+J\gamma_\Psi J^*)^{1/2} k h^{-1} k^* (1+J\gamma_\Psi J^*)^{1/2} \Big) \right]^{1/2} . 
\end{align*}
Using $\alpha_\Psi^* (1+J\gamma_\Psi J^*)^{-1}\alpha_\Psi\le \gamma_\Psi$ and $k h^{-1} k^* \le JhJ^*$, we get
\begin{align}
|\Tr(k^* \alpha_\Psi)| &\le \left[\Tr(h^{1/2} \gamma_\Psi h^{1/2}) \right]^{1/2}  \left[ \Tr( k h^{-1} k^*)+ \Tr(h^{1/2} \gamma_\Psi h^{1/2}) \right]^{1/2} \nn\\
& \le \Tr(h^{1/2} \gamma_\Psi h^{1/2}) + \frac{1}{2} \Tr( k h^{-1} k^*). \label{eq:|Tr-k*-alpha-Psi|-b}
\end{align}
Here in the last estimate we have used $\sqrt{x(x+y)}\le x+y/2$ for real numbers $x,y\ge 0$. 
Thus by definition \eqref{eq:def-H-gamma-alpha}, we get
$$
\langle \Psi, \bH \Psi \rangle = \Tr(h^{1/2} \gamma_\Psi h^{1/2}) + \Re \Tr(k^* \alpha_\Psi) \ge -\frac{1}{2} \Tr( k h^{-1} k^*)
$$
for $\Psi\in \mathcal{Q}$. Thus $\bH \ge -(1/2)\Tr( k h^{-1} k^*)$.  

The other inequality in Proposition \ref{lem:bounded-below-H} follows (using the additional assumption) from a similar argument as in \eqref{eq:|Tr-k*-alpha-Psi|-b}. This bound allows to conclude that the quadratic form $\bH$ is not only bounded below but also closed and thus its closure defines a self-adjoint operator.

 To prove Theorem \ref{thm:diag} we employ a connection between the bosonic diagonalization problem and its fermionic analogue. Such kind of connection has been known for a long time; see Araki \cite{Araki-68} for a heuristic  discussion. To be precise, we will use the following diagonalization result for fermionic block operators. 

\begin{theorem}[Diagonalization of fermionic block operators]\label{thm:diag-fermion}  Let $B$ be a self-adjoint operator on $\gH\oplus \gH^*$ such that $\cJ B \cJ=-B$ and such that $\dim \Ker(B)$ is either even (possibly $0$) or infinite. Then there exists a unitary operator $\cU$ on $\gH \oplus \gH^*$ such that $\cJ\cU \cJ=\cU$ and 
$$
\cU B \cU^* = \left( {\begin{array}{*{20}{c}}
  \xi & 0  \\ 
  0& -J \xi J^* 
\end{array}} \right)
$$
for some operator $\xi \ge0$ on $\gH$. Moreover, if $\Ker(B)=\{0\}$, then $\xi>0$. 
\end{theorem}

By applying Theorem \ref{thm:diag-fermion} to $B=\cA^{1/2}S\cA^{1/2}$, with $S$ given in \eqref{eq:def-S-J}, we can construct the Bogoliubov transformation $\cV$ in Theorem \ref{thm:diag} explicitly:
$$ \cV:= \cU |B|^{1/2} \cA^{-1/2}.$$
This explicit construction is similar to the one used by Simon, Chaturvedi and Srinivasan \cite{SimChaSri-1999} where they offered a simple proof of Williamsons' Theorem. The implementability of $\cV$ is proved using a detailed study of $ \cV^*\cV= \cA^{-1/2}|B|\cA^{-1/2}.$ 

To obtain the diagonalization result in Theorem \ref{thm:diag-Hamiltonian}, we recall from Proposition \ref{lem:bounded-below-H}, that the quadratic form $\bH$ defines a self-adjoint operator, still denoted by $\bH$, such that 
$$\inf \sigma(\bH) \ge -\frac{1}{2} \Tr(kh^{-1}k^*).$$
Let $\cV$ be as in Theorem \ref{thm:diag}. Let $\Psi$ be a normalized vector in $\mathcal{Q}$ defined in \eqref{eq:def-mcQ}. Consider $\Psi':=\bU_\cV^* \Psi$. It is straightforward to see that  
\begin{align}\label{eq:V*-Gamma-V} 
\left( {\begin{array}{*{20}{c}}
  \gamma_{\Psi'} & \alpha_{\Psi'}^*  \\ 
  \alpha_{\Psi'} & 1+ J \gamma_{\Psi'} J^* 
\end{array}} \right) = \cV^* \left( {\begin{array}{*{20}{c}}
  \gamma_{\Psi} & \alpha_{\Psi}^*  \\ 
  \alpha_{\Psi} & 1+ J \gamma_{\Psi} J^* 
\end{array}} \right) \cV.
\end{align}
Moreover, we have
\begin{align} \label{eq:V*-0001-V}
\cV^* 
\left( {\begin{array}{c c}
  0 & 0  \\ 
  0 & 1 
\end{array}} \right) \cV = \left( {\begin{array}{c c}
  X & Y^*  \\ 
  Y & 1+JXJ^* 
\end{array}} \right)
\end{align}
with $X = V^*V$ and $Y = JV^*JU = JU^*J^*V$. From \eqref{eq:V*-Gamma-V} and \eqref{eq:V*-0001-V}, we obtain 
\begin{align*}
\left( {\begin{array}{c c}
  \gamma_{\Psi'} - X & \alpha_{\Psi'}^* - Y^*  \\ 
  \alpha_{\Psi'}-Y & J (\gamma_{\Psi'}-X) J^* 
\end{array}} \right)  
&= \left( {\begin{array}{c c}
  \gamma_{\Psi'} & \alpha_{\Psi'}^*  \\ 
  \alpha_{\Psi'} & 1+ J \gamma_{\Psi'} J^* 
\end{array}} \right) -   \left( {\begin{array}{c c}
 X  & Y^*   \\ 
  Y & 1+ J X J^* \end{array}} \right) \\
& = \cV^* \left( {\begin{array}{c c}
  \gamma_{\Psi} & \alpha_{\Psi}^*  \\ 
  \alpha_{\Psi} & 1+ J \gamma_{\Psi} J^* 
\end{array}} \right) \cV - \cV^* 
\left( {\begin{array}{c c}
  0 & 0  \\ 
  0 & 1 
\end{array}} \right) \cV \\
&= \cV ^* \left( {\begin{array}{c c}
  \gamma_{\Psi} & \alpha_{\Psi}^*  \\ 
  \alpha_{\Psi} & J \gamma_{\Psi} J^* 
\end{array}} \right) \cV .
\end{align*}

Recall that $\gamma_\Psi$ and $\alpha_\Psi$ are finite-rank operators because $\Psi\in \mathcal{Q}$. Therefore,$\gamma_{\Psi'} - X$ and  $\alpha_{\Psi'}-Y$ are also finite-rank operators. Using the cyclicity of the trace we find that
\begin{align*} 
& \Tr(h^{1/2}(\gamma_{\Psi'}- X) h^{1/2}) + \Re \Tr(k^* (\alpha_{\Psi'}-Y)) \\
& = \frac{1}{2}  \Tr \left[ \cA  \left( {\begin{array}{c c}
  \gamma_{\Psi'} - X & \alpha_{\Psi'}^* - Y^*  \\ 
  \alpha_{\Psi'}-Y & J (\gamma_{\Psi'}-X) J^* 
\end{array}} \right)  \right] \\
& = \frac{1}{2}\Tr \left[ \cA  \cV^* \left( {\begin{array}{c c}
  \gamma_{\Psi} & \alpha_{\Psi}^* \\ 
  \alpha_{\Psi} & J \gamma_{\Psi}J^* 
\end{array}} \right) \cV  \right] 
 = \frac{1}{2}\Tr \left[ \cV \cA  \cV^* \left( {\begin{array}{c c}
  \gamma_{\Psi} & \alpha_{\Psi}^* \\ 
  \alpha_{\Psi} & J \gamma_{\Psi}J^* 
\end{array}} \right)  \right] \\
& = \frac{1}{2} \Tr \left[  \left( {\begin{array}{c c}
  \xi & 0 \\ 
  0 & J \xi  J^* 
\end{array}} \right) \left( {\begin{array}{c c}
  \gamma_{\Psi} & \alpha_{\Psi}^* \\ 
  \alpha_{\Psi} & J \gamma_{\Psi}J^* 
\end{array}} \right)  \right] = \Tr(\xi \gamma_\Psi) = \langle \Psi, \dGamma(\xi) \Psi \rangle.
\end{align*}
Thus by the quadratic form expression \eqref{eq:def-H-gamma-alpha}, we have
\begin{align*}
\langle \Psi, \bU_\cV \bH \bU_\cV^* \Psi \rangle &= \langle \bU_\cV^* \Psi, \bH \bU_\cV^* \Psi \rangle = \Tr(h^{1/2}\gamma_{\Psi'}h^{1/2})+ \Re \Tr(k^* \alpha_{\Psi'}) \\ 
&= \Tr(h^{1/2}(\gamma_{\Psi'}- X) h^{1/2}) + \Re \Tr(k^* (\alpha_{\Psi'}-Y)) \\
&\quad + \Tr(h^{1/2}Xh^{1/2})+ \Re \Tr(k^* Y) \\
&= \langle \Psi, \dGamma(\xi) \Psi \rangle + \Tr(h^{1/2}Xh^{1/2})+ \Re \Tr(k^* Y)
\end{align*}
for all $\Psi \in \mathcal{Q}$. A more detailed analysis shows that $h^{1/2}Xh^{1/2}$ and $k^* Y$ are trace class operators. Hence,  
$$ \bU_\cV \bH \bU_\cV^* = \dGamma(\xi) + \Tr(h^{1/2}X h^{1/2})+ \Re \Tr(k^* Y). $$
Since $\dGamma(\xi)$ has a unique ground state $|0\rangle$ with the ground state energy $0$, we conclude that $\bH$ has a unique ground state $\Psi_0=\bU_\cV^* |0\rangle$ with the ground state energy 
$$ \inf\sigma(\bH) = \Tr(h^{1/2}Xh^{1/2})+ \Re \Tr(k^*Y).$$
Finally, using \eqref{eq:V*-Gamma-V} and \eqref{eq:V*-0001-V} we find that $\gamma_{\Psi_0}=X$ and $\alpha_{\Psi_0}=Y$.

\section{Time-dependent Bogoliubov Hamiltonians}\label{sec:qf-dynamics}
 A crucial role in the theory of quadratic Hamiltonians is played by a class of states called \textit{quasi-free} states. In fact, ground states of the quadratic Hamiltonians are quasi-free states (see \cite[Theorem A.1]{LewNamSerSol-15}). Recall that a state $\Psi$ in Fock space $\cF(\gH)$ is called a quasi-free state if it has finite particle number expectation and satisfies Wick's Theorem: 
\begin{align}
&\langle \Psi, a^{\#}(f_{1}) a^{\#}(f_{2}) \cdots a^{\#}(f_{2n-1}) \Psi \rangle = 0,  \label{eq:Wick-1}\\
&\langle \Psi, a^{\#}(f_{1}) a^{\#}(f_{2}) \cdots a^{\#}(f_{2n})  \Psi \rangle = \sum_{\sigma\in P_{2n}} \prod_{j=1}^n \langle \Psi, a^{\#}(f_{\sigma(2j-1)}) a^{\#}(f_{\sigma(2j)}) \Psi \rangle \label{eq:Wick-2}
\end{align}
for all $n$ and for all $f_1,...,f_n \in \gH$, where $a^{\#}$ is either the creation or annihilation operator and $P_{2n}$ is the set of pairings,
$$
P_{2n} = \{\sigma \in S(2n)~ |~ \sigma(2j-1)<\min\{\sigma(2j),\sigma(2j+1)\}~\text{for~all}~ j\}.
$$

It is clear that if $\Psi$ is a quasi-free, then the projection $|\Psi\rangle \langle \Psi|$ is determined completely by its one-particle density matrices $\gamma_{\Psi}$ and $\alpha_{\Psi}$. States implemented by a Bogoliubov transformation, such as $\Psi_0=\mathbb{U}_{\cV}^* |0\rangle$ in Theorem \ref{thm:diag-Hamiltonian}, are called \textit{pure quasi-free} states. 

Since quasi-free states are so closely related to quadratic Hamiltonians, a natural question one can ask is whether this property (of being a quasi-free state) is preserved under time evolution generated by a Bogoliubov Hamiltonian. In case of time-independent quadratic Hamiltonians the answer is positive and the proof is rather straightforward. In the case of time-dependent Hamiltonians the same remains true, but the arguments are different and we will sketch them below (for more details see \cite{NamNap-15}, also see \cite{NamNap-17,NamNap-17a} for further applications).

Let  $\Phi(t)$ be a state in the Fock space  $\cF$ (which could, in general, depend on time). Assume $\Phi(t)$ satisfies the Bogoliubov equation
\begin{equation} \label{eq:Bogoliubov-equation}
i\partial_t \Phi(t) = \bH(t) \Phi(t), \quad  \Phi(t=0) = \Phi(0).
\end{equation}
Here $\bH(t)$ is a quadraric, time-dependent Hamiltonian in the Fock space: 
$$
\bH(t)= \dGamma(h(t)) + \frac12\iint_{\R^3\times\R^3}\Big(K_2(t,x,y)a^*_x a^*_y +\overline{K_2(t,x,y)}a_x a_y\Big)\d x\,\d y.
$$
 Let us focus on the special situation when $h(t)$ can be decomposed as
 $$h(t)=-\Delta +h_1(t).$$
For simplicity, assume furthermore that $h_1(t)$ and $K_2$ are bounded uniformly in time. We then have the following

\begin{proposition}[Bogoliubov equation] \label{lem:Bogoliubov-equation} (i) If $\Phi(0)$ belongs to the quadratic form domain $\mathcal{Q}(\dGamma (1-\Delta))$, then equation \eqref{eq:Bogoliubov-equation} has a unique global solution in $\mathcal{Q}(\dGamma (1-\Delta))$.
Moreover, the pair of density matrices $(\gamma_{\Phi(t)},\alpha_{\Phi(t)})$ is the unique solution to
\bq \label{eq:linear-Bog-dm} 
\left\{
\begin{aligned}
i\partial_t \gamma &= h \gamma - \gamma h + K_2 \alpha - \alpha^* K_2^*, \\
i\partial_t \alpha &= h \alpha + \alpha h^{\rm T} + K_2  + K_2 \gamma^{\rm T} + \gamma K_2,\\
\gamma(t&=0)=\gamma_{\Phi(0)}, \quad \alpha(t=0)  = \alpha_{\Phi(0)}.
\end{aligned}
\right.
\eq
(ii) If we assume further that $\Phi(0)$ is a quasi-free state in $\cF$, then $\Phi(t)$ is also a quasi-free state for all $t>0$.
\end{proposition}

Let us sketch the proof. Recall that $\gamma_{\Phi(t)}:\gH \to \gH$, $\alpha_{\Phi(t)}:\overline{\gH} \equiv \gH^* \to {\gH}$ are operators with kernels
$\gamma_{\Phi(t)}(x,y)=\langle \Phi(t), a_y^* a_x \Phi(t) \rangle$, $\alpha_{\Phi(t)}(x,y)=\langle \Phi(t), a_x a_y \Phi(t) \rangle
$ and $K_2: \overline{\gH} \equiv \gH^*\to \gH$ is an operator with kernel $K_2(t,x,y)$. 

For existence and uniqueness of $\Phi(t)$, we refer to \cite[Theorem 7]{LewNamSch-15}. To derive \eqref{eq:linear-Bog-dm}, we use \eqref{eq:Bogoliubov-equation} to compute 
\begin{align*}
&i\partial_t \gamma_{\Phi(t)}(x',y') = i\partial_t \langle \Phi(t), a_{y'}^* a_x \Phi(t) \rangle = \langle \Phi(t), [a_{y'}^* a_x , \bH(t)] \Phi(t) \rangle \\
&= \iint h(t,x,y) \Big(\delta (x'-x) \gamma_{\Phi(t)}(y,y') - \delta (y'-y) \gamma_{\Phi(t)}(x',x) \Big) \d x  \d y \\
& + \frac{1}{2} \iint k(t,x,y) \Big( \delta(x'-x) \alpha_{\Phi(t)}^*(y,y') + \delta (x'-y) \alpha_{\Phi(t)}^*(y',x) \Big) \d x  \d y \\
& - \frac{1}{2} \iint k^*(t,x,y) \Big( \delta(y'-y) \alpha_{\Phi(t)}(x,x') + \delta (y'-x) \alpha_{\Phi(t)}(y,x') \Big) \d x  \d y \\
& = \Big( h(t) \gamma_{\Phi(t)} - \gamma_{\Phi(t)} h(t) + K_2(t)\alpha^*_{\Phi(t)} - \alpha_{\Phi(t)} K_2^*(t) \Big)(x',y').
\end{align*}
This is the first equation in  \eqref{eq:linear-Bog-dm}. The second equation is proved similarly. 

Let us now discuss the second statement. We will write $(\gamma,\alpha)=(\gamma_{\Phi(t)}, \alpha_{\Phi(t)})$ for short. Let us introduce 
$$X:=\gamma+ \gamma^2 -\alpha \alpha^*, \quad Y:= \gamma \alpha - \alpha \gamma^{\rm T}.$$
It is a general fact (see, e.g., \cite[Lemma 8]{NamNap-15}) that $\Phi(t)$ is a quasi-free state if and only if $X(t)=0$ and $Y(t)=0$. In particular, we have $X(0)=0$ and $Y(0)=0$ by the assumption on $\Phi(0)$. Using \eqref{eq:linear-Bog-dm} it is straightforward to see that  
\begin{align*}
i\partial_t X &= h X- X h + K_2 Y^* - Y K_2^* ,\\
i\partial_t X^2 &= (i\partial_t X) X + X (i\partial_t X) = hX^2 - X^2 h + (K_2Y^* - Y K_2^*)X + X(K_2Y^* - Y K_2^*).
\end{align*}
Then we take the trace and use $\Tr(h X^2 - X^2 h)=0$ ($h X^2$ and $X^2 h$ may be not trace class but we can introduce a cut-off; see \cite{NamNap-15} for details). We find that
$$
\|X(t)\|_{\rm HS}^2  \le 4 \int_0^t \|K_2(s)\|\cdot \|X(s)\|_{\rm HS}\cdot \|Y(s)\|_{\rm HS} \,  \d s.
$$
We also obtain a similar bound for $\|Y(t)\|_{\rm HS}$. Summing these estimates and using the fact that $\|K_2(t)\|$ is bounded uniformly in time, we conclude by Gr\"onwall's inequality that $X(t)=0$, $Y(t)=0$ for all $t>0$. 

A similar argument can be used to show the uniqueness of solutions to \eqref{eq:linear-Bog-dm}.

\section{Time-dependent diagonalization problem}\label{sec:time-dep-diag}
In this section we would like to make a remark how equations \eqref{eq:Bogoliubov-equation} and \eqref{eq:linear-Bog-dm} are related to what can be called the \textit{time-dependent diagonalization problem}.

Assume we are given an effective (since quadratic rather than many-body, as explained in the introduction) evolution 
\begin{equation*} 
i\partial_t \Phi(t) = \bH(t) \Phi(t), \quad  \Phi(t=0) = \Phi(0).
\end{equation*}
Here $\bH(t)$ is a quadraric, time-dependent Hamiltonian in the Fock space. As stated in Proposition \ref{lem:Bogoliubov-equation}, if $\Phi(0)$ is a pure quasi-free state, then $\Phi(t)$ is also a pure quasi-free state for $t>0$.  Thus 
\bq \label{eq:time-dep-Bogo}
\Phi(t)=\mathbb{U}_{\cV}^*(t) |0\rangle
\eq
with, now, the Bogoliubov transformation being time-dependent. Inserting \eqref{eq:time-dep-Bogo} into \eqref{eq:Bogoliubov-equation} we obtain
\begin{equation*}
\left(i\partial_t \mathbb{U}_{\cV}^*(t)- \bH(t)\mathbb{U}_{\cV}^*(t)\right)  |0\rangle=0
\end{equation*} 
which implies
\bq \label{eq:time-dep-diag-Fock space}
\left(\mathbb{U}_{\cV}(t)(i\partial_t \mathbb{U}_{\cV}^*(t))-\mathbb{U}_{\cV}(t) \bH(t)\mathbb{U}_{\cV}^*(t)\right)  |0\rangle=0.  
\eq
It follows from \eqref{eq:time-dep-diag-Fock space} that 
\bq \label{eq:time-dep-diag-problem}
\mathbb{U}_{\cV}(t)(i\partial_t \mathbb{U}_{\cV}^*(t))-\mathbb{U}_{\cV}(t) \bH(t)\mathbb{U}_{\cV}^*(t)=\dGamma(\xi (t))  
\eq
for some operator $\xi(t): \h\to\h$. This argument can be inverted and one can ask the following question: given a (time-dependent) quadratic Hamiltonian $\bH (t)$, does there exist a unitarily implemented Bogoliubov transformation $\mathbb{U}_{\cV}(t)$ such that the operator
$$\mathbb{U}_{\cV}(t)(i\partial_t \mathbb{U}_{\cV}^*(t))-\mathbb{U}_{\cV}(t) \bH(t)\mathbb{U}_{\cV}^*(t)$$
is diagonal?  This question we shall call the \textit{time-dependent diagonalization problem}.

The above argument together with Proposition \ref{lem:Bogoliubov-equation} implies the following 
\begin{theorem}[Time-dependent diagonalization problem]
Let $\Phi(t)$ be a pure quasi-free state and and $\mathbb{U}_{\cV}^*(t)$ the corresponding  unitarily implemented Bogoliubov transformation such that $\Phi(t)=\mathbb{U}_{\cV}^*(t) |0\rangle$. Let  $(\gamma_{\Phi(t)},\alpha_{\Phi(t)})$ be the pair of density matrices associated to $\Phi(t)$. Let $\bH(t)$ be the quadratic operator defined in \eqref{eq:Bogoliubov-equation}. Then, under the assumptions of Proposition \ref{lem:Bogoliubov-equation},  we have that 
$$
\mathbb{U}_{\cV}(t)(i\partial_t \mathbb{U}_{\cV}^*(t))-\mathbb{U}_{\cV}(t) \bH(t)\mathbb{U}_{\cV}^*(t)=\dGamma(\xi (t))  
$$
for some operator $\xi(t): \h\to\h$ if and only if 
\bq \nonumber
\left\{
\begin{aligned}
i\partial_t \gamma &= h \gamma - \gamma h + K_2 \alpha - \alpha^* K_2^*, \\
i\partial_t \alpha &= h \alpha + \alpha h^{\rm T} + K_2  + K_2 \gamma^{\rm T} + \gamma K_2.
\end{aligned}
\right.
\eq
\end{theorem}

We would like to stress that formulating the time-dependent diagonalization problem in the language of quasi-free states yields a system of coupled linear equations for the reduced density matrices. Since solving linear equations is easier than non-linear ones, one could claim that this is the right approach. 

Indeed, a different approach would be to formulate this question in a more direct manner. As mentoined in the introduction, it is well known \cite{Berezin-66,BruDer-07} that pure quasi-free states can be represented in the explicit form
\begin{equation}\label{eq:exp-from-qfstate}
\mathbb{U}_{\cV_k}^*(t)|0\rangle = \exp\left ( i\chi_N(t) + \iint \left[ \overline{k(t,x,y)} a_x a_y - k(t,x,y) a^*_x a^*_y \right] \d x \d y \right) |0\rangle 
\end{equation}
 where $\chi_N(t) \in \R$ is a phase factor. One can now insert \eqref{eq:exp-from-qfstate} into the left hand side of \eqref{eq:time-dep-diag-problem} and demand that all terms involving $a^*_x a^*_y$ and $a_x a_y$ disappear. This condition will determine an equation for the kernel $k(t,x,y)$. To our knowledge, this approach was used for the first time in \cite{GriMacMar-10}. The resulting equation will be however highly non-linear. This can be easily seen from the relations 
\begin{equation}
\begin{aligned}
\mathbb{U}_{\cV_k}^*(t)a(f)\mathbb{U}_{\cV_k}(t)&=a(\cosh(2k)f)+a^*(\sinh(2k)\overline{f})\\
\mathbb{U}_{\cV_k}^*(t)a^*(f)\mathbb{U}_{\cV_k}(t)&=a^*(\cosh(2k)f)+a(\sinh(2k)\overline{f})
\end{aligned}
\end{equation}
for any $f\in \h$. Here $\cosh(2k)$ and $\sinh(2k)$ denote the linear operators on $\h$ given by
$$\cosh(2k):=\sum_{n\geq 0}\frac{1}{(2n)!}((2k)(\overline{2k}))^n,\, \sinh(2k):=\sum_{n\geq 0}\frac{1}{(2n+1)!}((2k)(2\overline{k}))^n (2k)$$
 where the products of $k$ have to be understood as products of operators (given by the kernels $k(t,x,y)$). In fact, this relation shows that the natural variables for the diagonalization problem are $\cosh(2k)$ and $\sinh(2k)$, but the resulting equations are still non-linear in those variables.  It has been then observed in \cite{GriMac-13} that when expressed in new variables, these equations  can be rewritten as linear equations that are equivalent to \eqref{eq:linear-Bog-dm}. 
 
 Adapting ideas from \cite{GriMac-13} to our setting, we shall now present an alternative derivation of \eqref{eq:Bogoliubov-equation}. To this end we will first rewrite the time-dependent diagonalization problem \eqref{eq:time-dep-diag-problem} on the classical level, that is before Weyl quantization. According to \eqref{eq:Wick-quant}, the classical counterpart of the term $\mathbb{U}_{\cV}(t) \bH(t)\mathbb{U}_{\cV}^*(t)$ is given by $\cV \cA \cV^*$ where $\cV$ and $\cA$ are defined in \eqref{eq:u-v-block} and 
 \eqref{eq:def-A-block} respectively (and, now, being time-dependent and with the off-diagonal term in $\cA$ being $K_2$ rather than $k$ according to the notation in \eqref{eq:Bogoliubov-equation}). Note that with $\mathbb{U}_{\cV_k}^*(t)$ given by \eqref{eq:exp-from-qfstate}, it follows from the formula for Weyl quantization \eqref{eq:Wick-quant} that
 $$\mathbb{U}_{\cV_k}^*(t)=e^{\bH_{\cK}}$$
 where 
 $$\cK= \left( {\begin{array}{c c}
  0 & -k  \\ 
  \overline{k} & 0 
\end{array}} \right).$$
 It follows that
$$\mathbb{U}_{\cV_k}(t)\partial_t \mathbb{U}_{\cV_k}^*(t)= e^{-\bH_{\cK}}\partial_t e^{\bH_{\cK}}.$$
Since  
\begin{equation*}
\begin{aligned}
e^{-\bH_{\cK}(t)}\partial_t e^{\bH_{\cK}(t)}=&\lim_{s\to 0}e^{-\bH_{\cK}(t)}\frac{1}{s}\left(e^{\bH_{\cK}(t+s)}-e^{\bH_{\cK}(t)}\right)\\
=&\lim_{s\to 0}\frac{1}{s}\int_0^1 d\lambda \frac{d}{d\lambda}\left(e^{-\bH_{\cK}(t)\lambda}e^{\bH_{\cK}(t+s)\lambda}\right) \\
=& \lim_{s\to 0}\int_0^1 d\lambda e^{-\bH_{\cK}(t)\lambda} \frac{-\bH_{\cK}(t)+\bH_{\cK}(t+s)}{s}e^{\bH_{\cK}(t+s)\lambda} \\
=& \int_0^1 d\lambda e^{-\bH_{\cK}(t)\lambda} \left(\partial_t \bH_{\cK}(t)\right)e^{\bH_{\cK}(t)\lambda}. 
\end{aligned}
\end{equation*}
using the so-called Lie formula we see that
$$e^{-\bH_{\cK}(t)}\partial_t e^{\bH_{\cK}(t)}=\int_0^1 d\lambda \left(\partial_t \bH_{\cK}(t)-[\bH_{\cK}(t) ,\partial_t \bH_{\cK}(t)]+\cdots \right).$$
Since the Weyl quantization is a Lie algebra isomorphism of antihermitian matrices we deduce that 
$$ e^{-\bH_{\cK}(t)}\partial_t e^{\bH_{\cK}(t)}= \bH_{e^{-\cK}\partial_t e^{\cK}}.$$
We already saw that 
$$e^{-\bH_{\cK}(t)} \bH_{\cA} e^{\bH_{\cK}(t)}= \bH_{e^{-\cK}\cA e^{\cK}}.$$
All this implies that on the classical level the time-dependent diagonalization problem is equivalent to 
$$\cV_{k}\widetilde{\cA} \cV^*_k$$ 
being block diagonal, with $\cV_{k}=e^{-\cK}$ and $\widetilde{\cA}$ is given \eqref{eq:def-A-block} where $h$ is replaced by $\widetilde{h}=i\partial_t+h$.

Now, recall that by general properties of Bogoliubov transformations, we have  $\cV_{k}^{-1}=S\cV_{k}^* S$. Since $\cV_k \widetilde{\cA} \cV_k^*=\cV_k\widetilde{\cA} S \cV_k^{-1}S$ is block diagonal, $\cV_k \widetilde{\cA} S \cV_k^{-1}$ is also block diagonal. Therefore, $[\cV_k \widetilde{\cA} S \cV_k^{-1},S]=0$, and hence $[\widetilde{\cA} S , \cV_k^{-1}S\cV_k ]=0$. Using $\cV_k^{-1}=S\cV_k^* S$ again we have
\bq
\cV_k^{-1}S\cV_k  = S\cV_k^{*}\cV_k 
=  \left( {\begin{array}{c c}
  1+2X_k & 2Y_k^*  \\ 
  - 2Y_k & -(1+ 2J X_k J^*) 
\end{array}} \right)
\eq
(recall \eqref{eq:V*-0001-V} for the definition of $X_k$ and $Y_k$). Therefore,
\begin{align*}
& 0 = [\cA S, \cV_k^{-1}S\cV_k] =  \left( {\begin{array}{c c}
  \widetilde{h} & -K_2^*   \\ 
  K_2  & - J \widetilde{h} J^* 
\end{array}} \right) \left( {\begin{array}{c c}
  1+2X_k & 2Y_k^*  \\ 
  - 2Y_k & -(1+ 2J X_k J^*) 
\end{array}} \right) \\
& \quad \quad\quad\quad\quad\quad\quad\quad - \left( {\begin{array}{c c}
  1+2X_k & 2Y_k^*  \\ 
  - 2Y_k & -(1+ 2J X_k J^*) 
\end{array}} \right) \left( {\begin{array}{c c}
  \widetilde{h} & -K_2^*   \\ 
  K_2  & - J \widetilde{h} J^* 
\end{array}} \right) \\
&= 2\left( {\begin{array}{*{20}{c}}
  \widetilde{h} X_k-X_k\widetilde{h}+K_2 ^*Y_k-Y_k^* K_2 & \widetilde{h} Y_k^*+ Y_k^* J \widetilde{h} J^* +  K_2^* JX_kJ^* + X_k K_2^* + K_2^* \\ 
  J \widetilde{h} J^* Y_k + Y_k \widetilde{h} + K_2 X_k + J X_k J^* K_2 + K_2 &  J \widetilde{h} X_k J^* - J X_k \widetilde{h} J^* + K_2 Y_k^* - Y_k K_2^* 
\end{array}} \right). 
\end{align*}
Finally, just as at the end of Section \ref{sec:proof-remarks}, we identify $\gamma$ with $X_k$ and $\alpha$ with $Y_k$. All this, together with the fact that in the setting described in Section \ref{sec:qf-dynamics} the operator $J$ simply yields a complex conjugation, leads to \eqref{eq:linear-Bog-dm}.

\end{document}